\documentclass[%
 aip,
 amsmath,amssymb,
 reprint,%
 numerical,
]{revtex4-1}

\usepackage{graphicx}
\usepackage{dcolumn}
\usepackage{bm}

\usepackage[utf8]{inputenc}
\usepackage[T1]{fontenc}
\usepackage{mathptmx}

\usepackage{latexsym}
\usepackage{amsmath}
\usepackage{epsfig}

\graphicspath{{./Figures/}}

\begin{document}

\preprint{AIP/CC-EOM}

\title[EOM-CC Approach for Intrinsic Losses]{Equation of motion coupled-cluster approach for intrinsic
losses in x-ray spectra}

\author{J. J. Rehr}
\affiliation{Dept.\ of Physics, Univ.\ of Washington Seattle, WA 98195, USA}
\author{F. D. Vila}
\affiliation{Dept.\ of Physics, Univ.\ of Washington Seattle, WA 98195, USA}
\author{J. J. Kas}
\affiliation{Dept.\ of Physics, Univ.\ of Washington Seattle, WA 98195, USA}
\author{N. Y. Hirshberg}
\affiliation{Dept.\ of Physics, Univ.\ of Washington Seattle, WA 98195, USA}

\author{K. Kowalski}
\affiliation{William R. Wiley Molecular Sciences Laboratory,
Battelle, Pacific Northwest National Laboratory, K8-91, PO Box 999,
Richland, WA, 99352, USA}
\author{B. Peng}
\affiliation{Advanced Computing, Mathematics, and Data Division, Battelle, Pacific Northwest National Laboratory, K8-91, PO Box 999, Richland, WA 99352, USA}



\date{\today}

\begin{abstract}
We present an equation of motion coupled cluster 
approach for calculating and understanding intrinsic inelastic
losses in core level x-ray absorption spectra (XAS).
The method is based on a factorization of the transition
amplitude in the time-domain, which leads to a
convolution of an effective one-body spectrum and
the core-hole spectral function.
The spectral function
characterizes these losses in terms of shake-up excitations and satellites,
and is calculated using a cumulant representation of the
core-hole Green's function that includes non-linear corrections.
The one-body spectrum also includes orthogonality corrections that
enhance the XAS at the edge. 
\end{abstract}

\maketitle



\section{\label{sec:intro}Introduction}

Calculations of x-ray absorption spectra (XAS) $\mu(\omega)$ from 
a deep core level of a many-electron system typically begin with
Fermi's golden rule
\begin{equation}
\mu(\omega) = \sum_F |\langle \Psi|D|\Psi_F\rangle|^2
 \delta(E_F-E_0 - \hbar\omega),
\end{equation}
where $|\Psi\rangle$ is the ground state with energy $E_0$,
$D=\Sigma_i d_i$ is the (dipole) interaction with
the x-ray field of frequency $\omega$, and the sum is
over the eigenstates $|\Psi_F\rangle$ of the many-body Hamiltonian $H$ with
energies $E_F$.   While full calculations are generally intractable,
the problem can be simplified in various ways.
For example, in the determinantal ${\Delta}$SCF approach, where
the initial and final many-body states are restricted to
single-Slater determinants,
the final states can be classified in terms of
successive single-, double-, and higher $n$-tuple
excitations.\cite{Liang2017,Liang2019} Alternatively with
Green's function methods, the summation over final states is
implicit.\cite{LeeVila,LeeBertsch12,Rehr09}
For molecular systems, for example, coupled-cluster (CC) Green's function
approaches have been developed both in
energy-space,\cite{PengKowalski2016,PengKowalski2018}
and in real
time.\cite{ChanWhite,KouliasLi19,arponen1983variational,kvaal2012ab,pedersen2019symplectic,nascimento2017simulation,nascimento2016linear,nascimento2019general,pigg2012time,sato2018communication}
While these developments generally focus on 
accurate calculations, relatively less attention has been
devoted to the analysis and understanding of many-body
effects in the spectra.

Our aim here is to to address this shortcoming.  To this end we introduce
a real-time equation of motion coupled-cluster (EOM-CC) approach
together with
a cumulant representation of the core-hole Green's function.
Cumulant techniques have been used increasingly to understand
correlation effects and exited states.\cite{Hedin99review,sky}
The approach provides an efficient method for calculations of inelastic
losses which simplifies their analysis and can be systematically improved.
  A key step in our approach is a factorization
of the XAS transition amplitude
 \begin{equation}
\mu(t) = \langle \Psi | D(0)  D(t) |\Psi\rangle = L(t) G_c(t)
\end{equation}
into an effective one-body transition amplitude $L(t)$ and
the core-hole Green's function $G_c(t)$.
This strategy is similar to that in the time-correlation approach
of Nozieres and de Dominicis\cite{ND} and Nozieres and Combescot (NC),\cite{NC}
for the edge-singularity problem. 
As a consequence the XAS is given by a convolution of
a one-electron cross-section $\mu_1(\omega)$ and the
core-hole spectral function $A_c(\omega)$ obtained from the Fourier transforms
of $L(t)$ and $G_c(t)$ respectively,
\begin{equation}
\mu(\omega) =  \int d\omega' \mu_1(\omega-\omega') A_c(\omega').
\end{equation}
The intrinsic inelastic losses due to the sudden creation of the 
core hole lead to shake-up effects characterized by 
satellite structure in the spectral function $A_c(\omega)$, 
and are directly related to x-ray photoemission spectra (XPS). 
The one-body spectrum $\mu_1(\omega)$ accounts for
edge-enhancement orthogonality corrections,
analogous to the prediction of Mahan.\cite{Mahan67}
However, the present approach ignores extrinsic
losses and interference which may likely decrease 
these effects.

In the remainder of this paper Sec.\ \ref{sec:eom} describes the EOM-CC approach
and Sec.\ \ref{sec:xray} the application to XAS. Finally, Secs. \ref{sec:calc}
and \ref{sec:summ} present prototypical results for a diatomic system
and a brief summary, respectively.

\section{\label{sec:eom}EOM-CC theory}


Intrinsic inelastic losses in XAS are implicit in the core-hole Green's
function $G_c(t)$ 
\begin{equation}
G_c(t) =  -i \langle \Psi_c|e^{i (H-E_0)  t} |\Psi_c\rangle \theta(t).
\end{equation}
Here $|\Psi_c\rangle = c_c|\Psi\rangle$ is the state of the
system at $t=0+$ with a core-hole in a deep level $|c\rangle$.
$H$ is the many-body Hamiltonian in the Hartree-Fock approximation,
and  $E_0$ is the ground state energy.
Our approach for calculating $G_c(t)$ is based on
the EOM-CC ansatz introduced by
Sch\"onhammer and Gunnarsson (SG),\cite{SG1978}
where $|\Psi\rangle$ and $|\Psi_c\rangle$ are taken to be single
Slater determinants.
The evolution of $|\Psi_c(t)\rangle$ is done
by transforming to an initial value problem,
and propagating according to the Schrodinger EOM
$i\, \partial |\Psi_c(t)\rangle/\partial t = H|\Psi_c(t)\rangle$, where 
$|\Psi_c(0)\rangle=c_c |\Psi\rangle$. The time-evolved state
$|\Psi_c(t)\rangle$ can be defined for any $t$ according to
a CC ansatz
\begin{equation}
|\Psi_c(t)\rangle \equiv N_c(t) e^{T_c(t)}|\Psi_c\rangle.
\end{equation}
For a non-interacting Hamiltonian, the CC ansatz for single-excitations
is justified by the Thouless theorem.\cite{Thouless}
Here $N_c(t)$ is a normalization factor and the time-dependent
CC operator $T_c(t)$ is defined in terms of single, double, etc.,
excitation creation operators 
$a^{\dagger}_n$, i.e., 
\begin{equation}
T_c(t)= \sum_{n} t_n(t) a^{\dagger}_n.
\end{equation}
For example, for the singles $n=(i,a)$ and $a_n^{\dagger} = c^{\dagger}_a c_i$; for the
doubles $n=i,j,a,b$ and $a_n^{\dagger}=c^{\dagger}_a c^{\dagger}_b c_j c_i$;
etc. Following the CC convention,
the indices $i,j$ refer to occupied, and $a,b,\dots$ to unoccupied
levels of the independent particle ground state.


Next by applying the Schrodinger EOM, left
multiplying by $e^{-T_c(t)}$,
and dividing by $N(t)$ yields the coupled EOM
\begin{equation}
\left[\frac{i\, \partial \ln N(t)}{\partial t} +
\frac{i\, \partial\, {T_c(t)} }{\partial t} \right]|\Psi_c\rangle =
\bar H(t)  |\Psi_c\rangle,
\end{equation}
where $\bar H(t) = e^{-T_c(t)} H(t) e^{T_c(t)}$ is
the similarity transformed Hamiltonian.
On applying successive commutation relations, the expansion of
$\bar H(t)$ terminates after two (four) terms for
single-particle (two-particle) operators.
Then left multiplying by $\langle \Psi|$ or $\langle n|=
\langle \Psi|a_n$
separates the EOM  as
\begin{eqnarray}
 {i\,  \partial \ln N_c(t)}/{\partial t} 
 &=&  \langle \Psi_c| \bar H(t)  |\Psi_c\rangle, \label{Neom} \\
 {i\, \partial\, t_n(t)}/{\partial t} &=&
\langle n | \bar H(t)|\Psi_c\rangle. \label{tneom}
\end{eqnarray}
As a result the core-hole Green's function is given 
by the time-dependent normalization factor
\begin{equation}
G_c(t)= -i N_c(t) \langle \Psi_c| e^{T_c(t)} |\Psi_c\rangle e^{-iE_0t} \theta(t)
= -i N_c(t)e^{-iE_0t} \theta(t).
\end{equation}
Moreover, Eq.\ (8) implies that $N_c(t)$ is a pure exponential,
so that $G_c(t)$ has a cumulant representation 
$G_c(t)=G_c^0(t) e^{C(t)}$, where 
$G_c^0(t)= -i e^{-i\epsilon_c t} \theta(t)$,  and
\begin{equation}
\label{NormCumulant}
C(t) =   -i \int_0^t dt'\,
\langle \Psi_c| (\bar H(t')-E'_0) |\Psi_c\rangle,
\end{equation}
where $E'_0=E_0-\epsilon_c$.  $C(t)$ can also be represented in
Landau form,\cite{Landau44} which simplifies the interpretation,
\begin{eqnarray}
C(t) &=& \int d\omega \frac{\beta(\omega)}{\omega^2}
 [e^{i\omega t} -i\omega t -1], \\
\beta(\omega) &=& \frac{1}{\pi} {\rm Re} \int_0^{\infty} dt\, e^{-i\omega t}
\frac{d}{dt} \langle \Psi_c|\bar H(t)|\Psi_c\rangle.
\end{eqnarray}
The cumulant kernel $\beta(\omega)$ accounts for the transfer of 
oscillator strength from the main peak to excitations at frequencies
$\omega$, and the initial conditions $C(0)=C'(0)=0$
guarantee its normalization and an invariant centroid.
Next we evaluate the cumulant in Eq.\ (\ref{NormCumulant}).
Our calculations are based on  an approximate Hartree-Fock Hamiltonian
for core-level XAS that assumes a single core-level localized
at an atomic site,\cite{langreth69} 
\begin{equation}
 H = \epsilon_c c_c^{\dagger} c_c + \sum_p \varepsilon_p c_p^{\dagger}c_p
   + \sum_{pq} v_{pq}^{cc} c_c c_c^{\dagger} c_p^{\dagger}c_q.
\end{equation}
Here $\epsilon_p$ are eigenstates of the initial state
one particle hamiltonian $h=\Sigma_p\epsilon_p c_p^{\dagger}c_p$ and
$h'=h+v$ is that for the final state in the presence of a core-hole
potential $v$, and $c_p^{\dagger}$ and $c_p$ are electron
creation and annihilation operators, respectively.
In order to illustrate the approach  here
we restrict the CC operator $T_c(t)$ to single-excitations
$T_c(t)=\Sigma_{a,i} t_{ai}(t) c_a^{\dagger}c_i$.
On applying the comutation relations for $[H,T]$ with Fermion
anticommutation properties, one obtains
\begin{equation}
\langle \Psi_c|(\bar H(t)-E'_0)|\Psi_c\rangle = 
 \sum_{ia} v_{ia}t_{ai}(t).
\end{equation}
From Eq.\ (\ref{tneom}), the coefficients $t_{ai}(t)$ obey a
first order non-linear differential equation\cite{SG1978}
\begin{eqnarray}
 i \frac{\partial t_{ai}}{\partial t} &=& \langle ai|\bar H(t)|\Psi_c\rangle
=  v_{ai} + \sum_b v_{ab}t_{bi}(t) \nonumber \\
  &-&  \sum_j t_{aj}(t)  v_{ji} -\sum_{bj}t_{aj}(t)v_{jb}t_{bi}(t).
\end{eqnarray}
This expression can  be interpreted perturbatively as a succession of
first order, second order, and third order terms in the off-diagonal matrix 
elements of $h'$.  The leading term in the cumulant
$v_{ia}t_{ia}^{1}(t)$ corresponds to linear response, and is second-order
in the core-hole potential $v$. The leading amplitude $t^{1}_{ia}(t)$
can be evaluated analytically to first order, yielding
$t^{1}_{ia}(t) = i ({v_{ai}}/{\omega_{ai}}) [ e^{i\omega_{ai} t} - 1 ]$,
where $\omega_{ai}=\epsilon_a -\epsilon_i$.
Inserting this result into Eq.\ (\ref{NormCumulant}),
we obtain an expression for $C(t)$ valid
to second order in $v$,
\begin{equation}
\beta(\omega) =  \sum_{ia} |v_{ia}|^2\delta(\omega-\omega_{ia}).
\end{equation}
 Higher order terms can be calculated systematically  and yield
non-linear (NL) corrections to the cumulant.\cite{Tzavala20,Mahan82}
For example, the third order term
can be obtained by inserting the 2nd order result above for $t^1$ above, and
so on.
For comparison we note that the core-hole spectral function can
also be obtained from a determinantal approach where
$G_c(t) = \det u_{ij}(t)$, $i,j =1,2\cdots N$, and
$u_{ij}(t) =\langle \phi_i| \phi_j(t)\rangle e^{-i\epsilon_i t}
= \langle i| e^{i (h'-\epsilon_i) t}|j\rangle$
are time-dependent overlap integrals.\cite{NC}

\section{\label{sec:xray}X-ray spectra}

The contribution to the XAS from a deep core level $|c\rangle$
is obtained using the time-correlation function with the factorization
$\mu(t)=L(t)G_c(t)$ in Eq.\ (2).
The core-hole Green's function $G_c$ is obtained
from Eq.\ (10-13).  Calculations of $L(t)$ can be done in various ways.
One is based on coupled EOM or equivalent integral
equations.\cite{langreth69,Grebennikov77,Agren01}
Another uses the time-evolution of one particle states,\cite{NC}
with the overlap integrals $u_{ij}(t)$ defined above.
Here we use a strategy similar to that of NC, except for the
replacement of the sums over $k$ with those for the
complete set of eigenstates $\kappa$ of $h'$.
Thus, defining the interaction operator for core transitions 
as $D = \Sigma_{\kappa} M_{c\kappa}c_{\kappa}^{\dagger}c_c$,
where $M_{c \kappa}=\langle c |d|\kappa \rangle$,
the one-body transition amplitude becomes
\begin{eqnarray}
L(t) &=& \sum_{\kappa,\kappa'} M^*_{c \kappa}M_{c \kappa'}
L_{\kappa,\kappa'}(t), \\
L_{\kappa,\kappa'}(t) &=& e^{i\epsilon_{\kappa}t} [u_{\kappa,\kappa'}(t) -
\sum^{occ}_{ij} u_{\kappa i}(t) u^{-1}_{ij}(t) u_{j\kappa'}(t) ].
\end{eqnarray}
The contribution to $L_{\kappa,\kappa'}(t)$ from the first term on the right
of Eq.~(19) leads to the independent particle transition amplitude
calculated in the presence of a core-hole 
$L_0(t)=\Sigma_{\kappa} |M_{c \kappa}|^2 \exp(i\epsilon_{\kappa} t),$
and is consistent with the final-state rule.\cite{Barth82}
The diagonal contributions $\kappa=\kappa'$ of the second term in (19)
contains the analog of a theta function $\theta(k_F-k)$ that
suppresses transitions to the occupied subspace $\kappa < k_F$.
The off-diagonal contributions to $L_{\kappa,\kappa'}(t)$ are dominated by
$\kappa$\ (or $\kappa'$) $> k_F$ and $\kappa'$\ (or $\kappa$) $ < k_F$,
respectively. The net result can be approximated 
by the compact expression
\begin{equation}
L(t) \approx  \sum_{\kappa} |\tilde M_{c\kappa}|^2 e^{i \epsilon_{\kappa} t},
\end{equation}
which is equivalent to that derived by Friedel,\cite{Friedel69}
and preserves the XAS sum-rule $\int d\omega \mu(\omega) = L(0)$.
Here $\tilde M_{c\kappa} =  \langle c | d \bar P |\kappa\rangle$,
where $\bar P = 1 - \Sigma^{N}_{i=1}|i\rangle\langle i| $
is the projection operator onto  the unoccupied valence levels of
the ground state. This approximation greatly simplifies the calculation of
the XAS, and we have verified that it agrees well with that using 
Eq.\ (19).
 The additional terms $-\Sigma_i\langle c|d|i\rangle \langle i|\kappa\rangle$
from $\bar P$ are called {\it replacement transitions}.
Physically, they serve to cancel transitions to
the occupied levels of the initial system.  To first order in
perturbation theory the
overlap $\langle i|\kappa\rangle \approx -v_{ik}/\omega_{ik}$ is negative for
an attractive core-hole potential and $\kappa > i$.  Thus they yield an
intrinsic edge enhancement factor $(1 + \chi_{\kappa})$ for each level $\kappa$
in the XAS where
$\chi_{\kappa} \approx -2 \Sigma_{i=1}^{N}
(M_{ci}/M_{c\kappa})\langle i|\kappa\rangle$.
While non-singular in molecular systems, this edge-enhancement effect
leads to the Mahan power-law singularity in metals\cite{Mahan67}
$\mu_1
\sim |(\epsilon -\epsilon_F)/\epsilon_F|^{-2\delta_l/\pi}$.
Finally the XAS in Eq.\ (3) is obtained by convolving $\mu_1(\omega)$
with $A_c(\omega)$.  For convenience we have shifted both
$\mu_1(\omega)$ and $A_c(\omega)$ by the core level
energy $\epsilon_c$, i.e. with $\omega=\epsilon-\epsilon_c$,
so that in the absence of interactions $\mu_1(\omega)$ agrees with
the independent particle XAS.  Formally the spectral function
represents the spectrum of shake-up excitations
\begin{equation}
A_c(\epsilon-\epsilon_c)
=\sum_{n} |S_n|^2 \delta(\epsilon+\epsilon_n),
\end{equation}
where $S_n = \langle \Psi_c|\Psi'_n\rangle$ is an $N-1$ body 
overlap integral and $\epsilon_n=E'_n-E_0$ is the net shake-up energy.
The behavior of $A_c(\omega)$  leads to
a significant reduction in the magnitude
of the XAS near the edge. In metals it leads to an Anderson power-law 
singularity\cite{ND} $[(\epsilon-\epsilon_F)/\epsilon_F)]^{\alpha}$.
This effect is opposite in sign, and thus competes with 
the enhancement from $\mu_1(\omega)$.




\begin{figure}[t]
\includegraphics[scale=0.32,clip,trim=0.0cm 1.2cm 2cm 3cm]{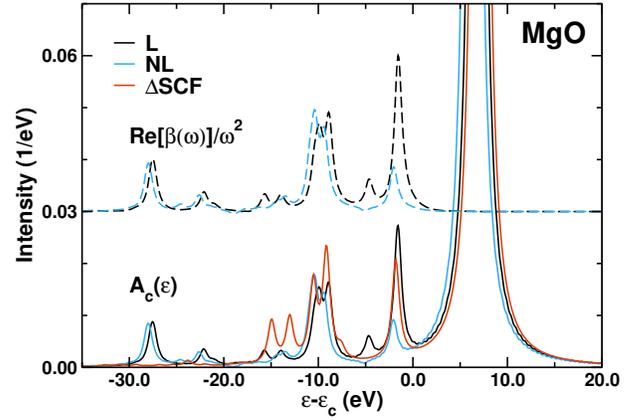}
\caption{\label{fig:beta_N2}
Cumulant kernel $\beta(\omega)$ (dashed lines)
compared to the the core spectral function $A_c(\omega)$ (full lines)
for the MgO molecule plotted vs energy relative to the bare
core hole energy $\epsilon_c =-1334.7 $ eV.
Both $A_c(\omega)$ and $\beta(\omega)$ were calculated with
either the linear (L) or non-linear (NL) approximations for $C(t)$;
the $\Delta$SCF calculation was obtained from the relation
$G_c(t)=\det u_{ij}(t)$ (see text).
Note that the non-linear terms introduce a small shift of
about -0.5 eV and reduce the satellite intensities.
}
\end{figure}
\begin{figure}[t]
\includegraphics[scale=0.32,clip,trim=0.0cm 1.2cm 2cm 2.8cm]{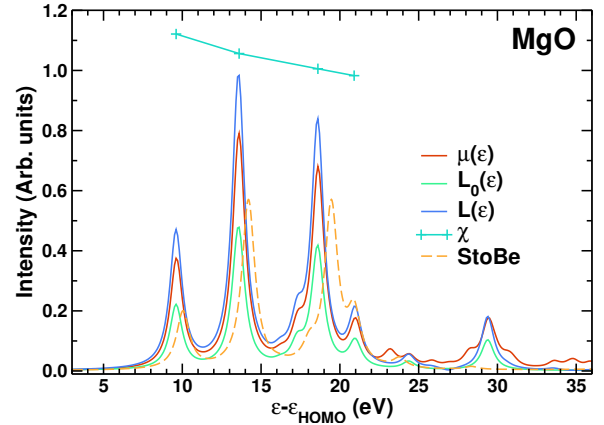}
\caption{\label{fig:xasmgo}
XAS $\mu(\omega)$
for MgO from the convolution in Eq.\ (3) compared to the
effective one-electron XAS $L(\epsilon)=\mu_1(\omega)$; 
the independent particle
XAS $L_0(\epsilon)=\mu_0(\omega)$ from the first term in Eq.\ (19), and 
from the independent electron code StoBe with a half-core-hole
approximation. Also shown is the edge enhancement
factor $\chi=L(\epsilon)/L_0(\epsilon) -1$.
All spectra are shown relative to the highest occupied molecular
orbital (HOMO) energy $\epsilon_{\rm HOMO}=$-8.4 eV.
}
\end{figure}

\section{\label{sec:calc}Calculations}

As an example, we present calculations for
a simple diatomic molecule MgO with bond distance 
1.749\AA.\cite{MgOdist} We use the cc-pVDZ basis set,\cite{prascher2011gaussian}
and the parameters and associated one-particle states
are obtained using a single-determinant Hartree-Fock reference.
Results for the cumulant kernel $\beta(\omega)$
and the spectral function $A_c(\omega)$ are shown in Fig.\ 1.
Note that the peaks in $\beta(\omega)$ correspond to the
inelastic losses in the spectral function, and for MgO are dominated by 
shake-up excitations just below the bare core hole energy peak.
These losses correspond to satellites visible in the XPS.
Remarkably these results show that the non-linear terms in the
cumulant are not large, so that the CC expansion converges rapidly
and linear response is a good approximation.  Consequently 
significant improvements in efficiency are possible with the CC-EOM
approach.
From the Landau form of the cumulant, the strength of the main
peak is given by the renormalization constant
$Z=\exp(-a) = 0.85\ (0.88)$ for the L (NL) cases, consistent with
direct integration over the main peak.
Here $a=\int d\omega\, \beta(\omega)/\omega^2 = 0.16\ (0.12)$
is the net satellite strength.  The quantity $\exp(-a)$
is also responsible for the amplitude reduction factor $S_0^2$
observed in the XAS fine structure.\cite{RehrStern78}
We have checked that  our results for $Z$ agree with those calculated
using the energy-space CC Green's
function approach.\cite{PengKowalski2016,PengKowalski2018}
Calculations of the XAS using the cc-pVDZ basis set are also
shown in Fig.\ \ref{fig:xasmgo}. For comparison, we also include the XAS
computed using the half core-hole approach in StoBe-deMon,\cite{stobe}
with a BE88PD86 exchange-correlation functional\cite{PhysRevA.38.3098,PhysRevB.33.8800} and the same basis set. Note that the StoBe-deMon results
agree well with the independent particle XAS $L_0$. The corrections to
the independent particle XAS in both $L(\epsilon)$ and $A_c(\epsilon)$
are substantial for MgO, but opposite in sign and dominated by
the edge enhancement factor $1+\chi$.
This effect can be traced to the magnitude of the core-valence
matrix elements $M_{ci}$ in $\chi_{\kappa}$.
Although $\chi$ exceeds unity for the bound-bound peaks near the edge,
this is likely an overestimate due to the neglect of extrinsic and 
interference effects.
On the other hand, satellite effects in the XAS are only weakly visible,
e.g., in the extra peaks between about 20 and 35 eV.

\section{\label{sec:summ}Summary and Conclusions}

We have  presented a real-time, EOM approach for calculations of
XAS including intrinsic losses, based on
the CC ansatz and a cumulant Green's function representation
of the core-hole spectral function.
 Although additional correlation is possible, for simplicity we
have limited our treatment here to single-determinant
wave-functions and the Hartree-Fock approximation.  The cumulant representation
facilitates both calculations and the interpretation of intrinsic
losses in the spectra.  A key step in our approach is a
time-domain factorization leading to a convolution formula
for the XAS in Eq.\ (2), in terms of the core-hole spectral function
and an effective one-particle spectrum.  These quantities account for
inelastic losses due to shake up excitations,
and edge enhancement corrections due to orthogonality, respectively. 
Though non-singular in molecular systems, both substantially
affect the XAS amplitude near threshold. While extrinsic
losses and interference terms due to the coupling of the
photo-electron to the core-hole are ignored in this treatment,
these effects are opposite in sign and tend to cancel.
 Remarkably the calculation of the cumulant converges rapidly,
yielding good results even for the second-order or linear-response
approximation.
The nature of the CC-EOM cumulant is analogous to that encountered in other
theoretical treatments, e.g., using the linked-cluster
theorem, field-theoretic methods, or the quasi-boson
approximation.\cite{langreth70,ND,Hedin99review}
In condensed matter the cumulant kernel $\beta(\omega)$
is directly related to the loss function, and characterizes
excitations such as density fluctuations due to the suddence
appearance of the core-hole.\cite{langreth70,KasRehr17}
 Many extensions of the methodology introduced here are possible. 
For example, the treatment of emission spectra is directly analogous
to that for XAS.\cite{NC}
A more extensive treatment including the extension to higher order
CCSD excitations will be presented elsewhere.\cite{VilaRehr}

\begin{acknowledgments}
We thank X. Li, D. Prendergast, K. Sch\"onhammer, and M. Tzavala, 
for comments, and D. Adkins and D. Share for encouragement.
 This work was supported by the Computational
Chemical Sciences Program of the U.S. Department
of Energy, Office of Science, BES, Chemical Sciences, Geosciences
and Biosciences Division in the Center for Scalable
and Predictive methods for Excitations and Correlated phenomena
(SPEC) at PNNL. One of us (NYH) acknowledges support from by the
NSF REU Program in summer 2019.
\end{acknowledgments}

\bibliographystyle{apsrev}
\bibliography{references}

\begin{thebibliography}{41}
\expandafter\ifx\csname natexlab\endcsname\relax\def\natexlab#1{#1}\fi
\expandafter\ifx\csname bibnamefont\endcsname\relax
  \def\bibnamefont#1{#1}\fi
\expandafter\ifx\csname bibfnamefont\endcsname\relax
  \def\bibfnamefont#1{#1}\fi
\expandafter\ifx\csname citenamefont\endcsname\relax
  \def\citenamefont#1{#1}\fi
\expandafter\ifx\csname url\endcsname\relax
  \def\url#1{\texttt{#1}}\fi
\expandafter\ifx\csname urlprefix\endcsname\relax\def\urlprefix{URL }\fi
\providecommand{\bibinfo}[2]{#2}
\providecommand{\eprint}[2][]{\url{#2}}

\bibitem[{\citenamefont{Liang et~al.}(2017)\citenamefont{Liang, Vinson,
  Pemmaraju, Drisdell, Shirley, and Prendergast}}]{Liang2017}
\bibinfo{author}{\bibfnamefont{Y.}~\bibnamefont{Liang}},
  \bibinfo{author}{\bibfnamefont{J.}~\bibnamefont{Vinson}},
  \bibinfo{author}{\bibfnamefont{S.}~\bibnamefont{Pemmaraju}},
  \bibinfo{author}{\bibfnamefont{W.~S.} \bibnamefont{Drisdell}},
  \bibinfo{author}{\bibfnamefont{E.~L.} \bibnamefont{Shirley}},
  \bibnamefont{and}
  \bibinfo{author}{\bibfnamefont{D.}~\bibnamefont{Prendergast}},
  \bibinfo{journal}{Phys. Rev. Lett.} \textbf{\bibinfo{volume}{118}},
  \bibinfo{pages}{096402} (\bibinfo{year}{2017}).

\bibitem[{\citenamefont{Liang and Prendergast}(2019)}]{Liang2019}
\bibinfo{author}{\bibfnamefont{Y.}~\bibnamefont{Liang}} \bibnamefont{and}
  \bibinfo{author}{\bibfnamefont{D.}~\bibnamefont{Prendergast}}
  (\bibinfo{year}{2019}), \bibinfo{note}{arXiv:1905.00542v1}.

\bibitem[{\citenamefont{Lee et~al.}(2012)\citenamefont{Lee, Vila, and
  Rehr}}]{LeeVila}
\bibinfo{author}{\bibfnamefont{A.~J.} \bibnamefont{Lee}},
  \bibinfo{author}{\bibfnamefont{F.~D.} \bibnamefont{Vila}}, \bibnamefont{and}
  \bibinfo{author}{\bibfnamefont{J.~J.} \bibnamefont{Rehr}},
  \bibinfo{journal}{Phys. Rev. B} \textbf{\bibinfo{volume}{86}},
  \bibinfo{pages}{115107} (\bibinfo{year}{2012}).

\bibitem[{\citenamefont{Bertsch and Lee}(2014)}]{LeeBertsch12}
\bibinfo{author}{\bibfnamefont{G.~F.} \bibnamefont{Bertsch}} \bibnamefont{and}
  \bibinfo{author}{\bibfnamefont{A.~J.} \bibnamefont{Lee}},
  \bibinfo{journal}{Phys. Rev. B} \textbf{\bibinfo{volume}{89}},
  \bibinfo{pages}{075135} (\bibinfo{year}{2014}).

\bibitem[{\citenamefont{Rehr et~al.}(2009)\citenamefont{Rehr, Kas, Prange,
  Sorini, Takimoto, and Vila}}]{Rehr09}
\bibinfo{author}{\bibfnamefont{J.~J.} \bibnamefont{Rehr}},
  \bibinfo{author}{\bibfnamefont{J.~J.} \bibnamefont{Kas}},
  \bibinfo{author}{\bibfnamefont{M.~P.} \bibnamefont{Prange}},
  \bibinfo{author}{\bibfnamefont{A.~P.} \bibnamefont{Sorini}},
  \bibinfo{author}{\bibfnamefont{Y.}~\bibnamefont{Takimoto}}, \bibnamefont{and}
  \bibinfo{author}{\bibfnamefont{F.}~\bibnamefont{Vila}},
  \bibinfo{journal}{Comptes Rendus Physique} \textbf{\bibinfo{volume}{10}},
  \bibinfo{pages}{548} (\bibinfo{year}{2009}).

\bibitem[{\citenamefont{Peng and Kowalski}(2016)}]{PengKowalski2016}
\bibinfo{author}{\bibfnamefont{B.}~\bibnamefont{Peng}} \bibnamefont{and}
  \bibinfo{author}{\bibfnamefont{K.}~\bibnamefont{Kowalski}},
  \bibinfo{journal}{Phys. Rev. A} \textbf{\bibinfo{volume}{94}},
  \bibinfo{pages}{062512} (\bibinfo{year}{2016}).

\bibitem[{\citenamefont{Peng and Kowalski}(2018)}]{PengKowalski2018}
\bibinfo{author}{\bibfnamefont{B.}~\bibnamefont{Peng}} \bibnamefont{and}
  \bibinfo{author}{\bibfnamefont{K.}~\bibnamefont{Kowalski}},
  \bibinfo{journal}{J. Chem. Theory Comput.} \textbf{\bibinfo{volume}{14}},
  \bibinfo{pages}{4335} (\bibinfo{year}{2018}).

\bibitem[{\citenamefont{White and Chan}(2018)}]{ChanWhite}
\bibinfo{author}{\bibfnamefont{A.~F.} \bibnamefont{White}} \bibnamefont{and}
  \bibinfo{author}{\bibfnamefont{G.~K.-L.} \bibnamefont{Chan}},
  \bibinfo{journal}{J. Chem. Theory. Comput.} \textbf{\bibinfo{volume}{14}},
  \bibinfo{pages}{5690} (\bibinfo{year}{2018}).

\bibitem[{\citenamefont{Koulias et~al.}(2019)\citenamefont{Koulias,
  Williams-Young, Nascimento, DePrince, and Li}}]{KouliasLi19}
\bibinfo{author}{\bibfnamefont{L.~N.} \bibnamefont{Koulias}},
  \bibinfo{author}{\bibfnamefont{D.~B.} \bibnamefont{Williams-Young}},
  \bibinfo{author}{\bibfnamefont{D.~R.} \bibnamefont{Nascimento}},
  \bibinfo{author}{\bibfnamefont{A.~E.} \bibnamefont{DePrince},
  \bibfnamefont{III}}, \bibnamefont{and}
  \bibinfo{author}{\bibfnamefont{X.}~\bibnamefont{Li}}, \bibinfo{journal}{J.
  Chem. Theory Comput.} \textbf{\bibinfo{volume}{15}}, \bibinfo{pages}{6617}
  (\bibinfo{year}{2019}).

\bibitem[{\citenamefont{Arponen}(1983)}]{arponen1983variational}
\bibinfo{author}{\bibfnamefont{J.}~\bibnamefont{Arponen}},
  \bibinfo{journal}{Ann. Phys.} \textbf{\bibinfo{volume}{151}},
  \bibinfo{pages}{311} (\bibinfo{year}{1983}).

\bibitem[{\citenamefont{Kvaal}(2012)}]{kvaal2012ab}
\bibinfo{author}{\bibfnamefont{S.}~\bibnamefont{Kvaal}}, \bibinfo{journal}{J.
  Chem. Phys.} \textbf{\bibinfo{volume}{136}}, \bibinfo{pages}{194109}
  (\bibinfo{year}{2012}).

\bibitem[{\citenamefont{Pedersen and Kvaal}(2019)}]{pedersen2019symplectic}
\bibinfo{author}{\bibfnamefont{T.~B.} \bibnamefont{Pedersen}} \bibnamefont{and}
  \bibinfo{author}{\bibfnamefont{S.}~\bibnamefont{Kvaal}}, \bibinfo{journal}{J.
  Chem. Phys.} \textbf{\bibinfo{volume}{150}}, \bibinfo{pages}{144106}
  (\bibinfo{year}{2019}).

\bibitem[{\citenamefont{Nascimento and
  DePrince~III}(2017)}]{nascimento2017simulation}
\bibinfo{author}{\bibfnamefont{D.~R.} \bibnamefont{Nascimento}}
  \bibnamefont{and} \bibinfo{author}{\bibfnamefont{A.~E.}
  \bibnamefont{DePrince~III}}, \bibinfo{journal}{J. Phys. Chem. Lett.}
  \textbf{\bibinfo{volume}{8}}, \bibinfo{pages}{2951} (\bibinfo{year}{2017}).

\bibitem[{\citenamefont{Nascimento and
  DePrince~III}(2016)}]{nascimento2016linear}
\bibinfo{author}{\bibfnamefont{D.~R.} \bibnamefont{Nascimento}}
  \bibnamefont{and} \bibinfo{author}{\bibfnamefont{A.~E.}
  \bibnamefont{DePrince~III}}, \bibinfo{journal}{J. Chem. Theory Comput.}
  \textbf{\bibinfo{volume}{12}}, \bibinfo{pages}{5834} (\bibinfo{year}{2016}).

\bibitem[{\citenamefont{Nascimento and
  DePrince~III}(2019)}]{nascimento2019general}
\bibinfo{author}{\bibfnamefont{D.~R.} \bibnamefont{Nascimento}}
  \bibnamefont{and} \bibinfo{author}{\bibfnamefont{A.~E.}
  \bibnamefont{DePrince~III}}, \bibinfo{journal}{J. Chem. Phys.}
  \textbf{\bibinfo{volume}{151}}, \bibinfo{pages}{204107}
  (\bibinfo{year}{2019}).

\bibitem[{\citenamefont{Pigg et~al.}(2012)\citenamefont{Pigg, Hagen, Nam, and
  Papenbrock}}]{pigg2012time}
\bibinfo{author}{\bibfnamefont{D.~A.} \bibnamefont{Pigg}},
  \bibinfo{author}{\bibfnamefont{G.}~\bibnamefont{Hagen}},
  \bibinfo{author}{\bibfnamefont{H.}~\bibnamefont{Nam}}, \bibnamefont{and}
  \bibinfo{author}{\bibfnamefont{T.}~\bibnamefont{Papenbrock}},
  \bibinfo{journal}{Phys. Rev. C} \textbf{\bibinfo{volume}{86}},
  \bibinfo{pages}{014308} (\bibinfo{year}{2012}).

\bibitem[{\citenamefont{Sato et~al.}(2018)\citenamefont{Sato, Pathak, Orimo,
  and Ishikawa}}]{sato2018communication}
\bibinfo{author}{\bibfnamefont{T.}~\bibnamefont{Sato}},
  \bibinfo{author}{\bibfnamefont{H.}~\bibnamefont{Pathak}},
  \bibinfo{author}{\bibfnamefont{Y.}~\bibnamefont{Orimo}}, \bibnamefont{and}
  \bibinfo{author}{\bibfnamefont{K.~L.} \bibnamefont{Ishikawa}},
  \bibinfo{journal}{J. Chem. Phys.} \textbf{\bibinfo{volume}{148}},
  \bibinfo{pages}{051101} (\bibinfo{year}{2018}).

\bibitem[{\citenamefont{Hedin}(1999)}]{Hedin99review}
\bibinfo{author}{\bibfnamefont{L.}~\bibnamefont{Hedin}}, \bibinfo{journal}{J.
  Phys.: Condens. Matter} \textbf{\bibinfo{volume}{11}}, \bibinfo{pages}{R489}
  (\bibinfo{year}{1999}).

\bibitem[{\citenamefont{Zhou et~al.}(2015)\citenamefont{Zhou, Kas, Sponza,
  Reshetnyak, Guzzo, Giorgetti, Gatti, Sottile, Rehr, and Reining}}]{sky}
\bibinfo{author}{\bibfnamefont{J.}~\bibnamefont{Zhou}},
  \bibinfo{author}{\bibfnamefont{J.}~\bibnamefont{Kas}},
  \bibinfo{author}{\bibfnamefont{L.}~\bibnamefont{Sponza}},
  \bibinfo{author}{\bibfnamefont{I.}~\bibnamefont{Reshetnyak}},
  \bibinfo{author}{\bibfnamefont{M.}~\bibnamefont{Guzzo}},
  \bibinfo{author}{\bibfnamefont{C.}~\bibnamefont{Giorgetti}},
  \bibinfo{author}{\bibfnamefont{M.}~\bibnamefont{Gatti}},
  \bibinfo{author}{\bibfnamefont{F.}~\bibnamefont{Sottile}},
  \bibinfo{author}{\bibfnamefont{J.}~\bibnamefont{Rehr}}, \bibnamefont{and}
  \bibinfo{author}{\bibfnamefont{L.}~\bibnamefont{Reining}},
  \bibinfo{journal}{J. Chem. Phys.} \textbf{\bibinfo{volume}{143}},
  \bibinfo{pages}{184109} (\bibinfo{year}{2015}).

\bibitem[{\citenamefont{Nozieres and Dominicis}(1969)}]{ND}
\bibinfo{author}{\bibfnamefont{P.}~\bibnamefont{Nozieres}} \bibnamefont{and}
  \bibinfo{author}{\bibfnamefont{C.~D.} \bibnamefont{Dominicis}},
  \bibinfo{journal}{Phys. Rev.} \textbf{\bibinfo{volume}{178}},
  \bibinfo{pages}{1097} (\bibinfo{year}{1969}).

\bibitem[{\citenamefont{Nozieres and Combescot}(1971)}]{NC}
\bibinfo{author}{\bibfnamefont{P.}~\bibnamefont{Nozieres}} \bibnamefont{and}
  \bibinfo{author}{\bibfnamefont{M.}~\bibnamefont{Combescot}},
  \bibinfo{journal}{J. de Physique} \textbf{\bibinfo{volume}{32}},
  \bibinfo{pages}{11} (\bibinfo{year}{1971}).

\bibitem[{\citenamefont{Mahan}(1967)}]{Mahan67}
\bibinfo{author}{\bibfnamefont{G.~D.} \bibnamefont{Mahan}},
  \bibinfo{journal}{Phys. Rev.} \textbf{\bibinfo{volume}{163}},
  \bibinfo{pages}{612} (\bibinfo{year}{1967}).

\bibitem[{\citenamefont{Sch\"onhammer and Gunnarsson}(1978)}]{SG1978}
\bibinfo{author}{\bibfnamefont{K.}~\bibnamefont{Sch\"onhammer}}
  \bibnamefont{and}
  \bibinfo{author}{\bibfnamefont{O.}~\bibnamefont{Gunnarsson}},
  \bibinfo{journal}{Phys. Rev. B} \textbf{\bibinfo{volume}{18}},
  \bibinfo{pages}{6606} (\bibinfo{year}{1978}).

\bibitem[{\citenamefont{Thouless}(1961)}]{Thouless}
\bibinfo{author}{\bibfnamefont{D.}~\bibnamefont{Thouless}},
  \emph{\bibinfo{title}{The Quantum Mechanics of Many-Body Systems}}
  (\bibinfo{publisher}{Academic, New York}, \bibinfo{year}{1961}).

\bibitem[{\citenamefont{Landau}(1944)}]{Landau44}
\bibinfo{author}{\bibfnamefont{L.}~\bibnamefont{Landau}}, \bibinfo{journal}{J.
  Phys. USSR} \textbf{\bibinfo{volume}{8}}, \bibinfo{pages}{201}
  (\bibinfo{year}{1944}).

\bibitem[{\citenamefont{Langreth}(1969)}]{langreth69}
\bibinfo{author}{\bibfnamefont{D.~C.} \bibnamefont{Langreth}},
  \bibinfo{journal}{Phys. Rev.} \textbf{\bibinfo{volume}{182}},
  \bibinfo{pages}{973} (\bibinfo{year}{1969}).

\bibitem[{\citenamefont{Tzavala et~al.}(2020)\citenamefont{Tzavala, Kas,
  Reining, and Rehr}}]{Tzavala20}
\bibinfo{author}{\bibfnamefont{M.}~\bibnamefont{Tzavala}},
  \bibinfo{author}{\bibfnamefont{J.~J.} \bibnamefont{Kas}},
  \bibinfo{author}{\bibfnamefont{L.}~\bibnamefont{Reining}}, \bibnamefont{and}
  \bibinfo{author}{\bibfnamefont{J.~J.} \bibnamefont{Rehr}},
  \bibinfo{journal}{Unpublished}  (\bibinfo{year}{2020}).

\bibitem[{\citenamefont{Mahan}(1982)}]{Mahan82}
\bibinfo{author}{\bibfnamefont{G.~D.} \bibnamefont{Mahan}},
  \bibinfo{journal}{Phys. Rev. B} \textbf{\bibinfo{volume}{25}},
  \bibinfo{pages}{5021} (\bibinfo{year}{1982}).

\bibitem[{\citenamefont{Grebennikov et~al.}(1977)\citenamefont{Grebennikov,
  Babanov, and Sokolov}}]{Grebennikov77}
\bibinfo{author}{\bibfnamefont{V.}~\bibnamefont{Grebennikov}},
  \bibinfo{author}{\bibfnamefont{Y.}~\bibnamefont{Babanov}}, \bibnamefont{and}
  \bibinfo{author}{\bibfnamefont{O.}~\bibnamefont{Sokolov}},
  \bibinfo{journal}{Phys. Stat. Sol. (b)} \textbf{\bibinfo{volume}{79}},
  \bibinfo{pages}{423} (\bibinfo{year}{1977}).

\bibitem[{\citenamefont{Privalov et~al.}(2001)\citenamefont{Privalov,
  Gel'mukhanov, and \AA{}gren}}]{Agren01}
\bibinfo{author}{\bibfnamefont{T.}~\bibnamefont{Privalov}},
  \bibinfo{author}{\bibfnamefont{F.}~\bibnamefont{Gel'mukhanov}},
  \bibnamefont{and}
  \bibinfo{author}{\bibfnamefont{H.}~\bibnamefont{\AA{}gren}},
  \bibinfo{journal}{Phys. Rev. B} \textbf{\bibinfo{volume}{64}},
  \bibinfo{pages}{165115} (\bibinfo{year}{2001}).

\bibitem[{\citenamefont{von Barth and Grossmann}(1982)}]{Barth82}
\bibinfo{author}{\bibfnamefont{U.}~\bibnamefont{von Barth}} \bibnamefont{and}
  \bibinfo{author}{\bibfnamefont{G.}~\bibnamefont{Grossmann}},
  \bibinfo{journal}{Phys. Rev. B} \textbf{\bibinfo{volume}{25}},
  \bibinfo{pages}{5150} (\bibinfo{year}{1982}).

\bibitem[{\citenamefont{Friedel}(1969)}]{Friedel69}
\bibinfo{author}{\bibfnamefont{J.}~\bibnamefont{Friedel}},
  \bibinfo{journal}{Comments Solid State Phys.} \textbf{\bibinfo{volume}{2}},
  \bibinfo{pages}{21} (\bibinfo{year}{1969}).

\bibitem[{\citenamefont{Huber and Herzberg}(retrieved January 22,
  2020)}]{MgOdist}
\bibinfo{author}{\bibfnamefont{K.~P.} \bibnamefont{Huber}} \bibnamefont{and}
  \bibinfo{author}{\bibfnamefont{G.~H.} \bibnamefont{Herzberg}}, in
  \emph{\bibinfo{booktitle}{Constants of Diatomic Molecules}}, edited by
  \bibinfo{editor}{\bibfnamefont{P.}~\bibnamefont{Linstrom}} \bibnamefont{and}
  \bibinfo{editor}{\bibfnamefont{W.}~\bibnamefont{Mallard}}
  (\bibinfo{publisher}{National Institute of Standards and Technology,
  Gaithersburg MD, 20899}, \bibinfo{year}{retrieved January 22, 2020}),
  vol.~\bibinfo{volume}{69} of \emph{\bibinfo{series}{NIST Chemistry WebBook,
  NIST Standard Reference Database}}.

\bibitem[{\citenamefont{Prascher et~al.}(2011)\citenamefont{Prascher, Woon,
  Peterson, Dunning, and Wilson}}]{prascher2011gaussian}
\bibinfo{author}{\bibfnamefont{B.~P.} \bibnamefont{Prascher}},
  \bibinfo{author}{\bibfnamefont{D.~E.} \bibnamefont{Woon}},
  \bibinfo{author}{\bibfnamefont{K.~A.} \bibnamefont{Peterson}},
  \bibinfo{author}{\bibfnamefont{T.~H.} \bibnamefont{Dunning}},
  \bibnamefont{and} \bibinfo{author}{\bibfnamefont{A.~K.}
  \bibnamefont{Wilson}}, \bibinfo{journal}{Theoretical Chemistry Accounts}
  \textbf{\bibinfo{volume}{128}}, \bibinfo{pages}{69} (\bibinfo{year}{2011}).

\bibitem[{\citenamefont{Rehr et~al.}(1978)\citenamefont{Rehr, Stern, Martin,
  and Davidson}}]{RehrStern78}
\bibinfo{author}{\bibfnamefont{J.~J.} \bibnamefont{Rehr}},
  \bibinfo{author}{\bibfnamefont{E.~A.} \bibnamefont{Stern}},
  \bibinfo{author}{\bibfnamefont{R.~L.} \bibnamefont{Martin}},
  \bibnamefont{and} \bibinfo{author}{\bibfnamefont{E.~R.}
  \bibnamefont{Davidson}}, \bibinfo{journal}{Phys. Rev. B}
  \textbf{\bibinfo{volume}{17}}, \bibinfo{pages}{560} (\bibinfo{year}{1978}).

\bibitem[{\citenamefont{Triguero et~al.}(1998)\citenamefont{Triguero,
  Pettersson, and \AA{}gren}}]{stobe}
\bibinfo{author}{\bibfnamefont{L.}~\bibnamefont{Triguero}},
  \bibinfo{author}{\bibfnamefont{L.}~\bibnamefont{Pettersson}},
  \bibnamefont{and}
  \bibinfo{author}{\bibfnamefont{H.}~\bibnamefont{\AA{}gren}},
  \bibinfo{journal}{J. Phys. Chem. A} \textbf{\bibinfo{volume}{102}},
  \bibinfo{pages}{10599} (\bibinfo{year}{1998}).

\bibitem[{\citenamefont{Becke}(1988)}]{PhysRevA.38.3098}
\bibinfo{author}{\bibfnamefont{A.~D.} \bibnamefont{Becke}},
  \bibinfo{journal}{Phys. Rev. A} \textbf{\bibinfo{volume}{38}},
  \bibinfo{pages}{3098} (\bibinfo{year}{1988}).

\bibitem[{\citenamefont{Perdew and Yue}(1986)}]{PhysRevB.33.8800}
\bibinfo{author}{\bibfnamefont{J.~P.} \bibnamefont{Perdew}} \bibnamefont{and}
  \bibinfo{author}{\bibfnamefont{W.}~\bibnamefont{Yue}},
  \bibinfo{journal}{Phys. Rev. B} \textbf{\bibinfo{volume}{33}},
  \bibinfo{pages}{8800} (\bibinfo{year}{1986}).

\bibitem[{\citenamefont{Langreth}(1970)}]{langreth70}
\bibinfo{author}{\bibfnamefont{D.~C.} \bibnamefont{Langreth}},
  \bibinfo{journal}{Phys. Rev. B} \textbf{\bibinfo{volume}{1}},
  \bibinfo{pages}{471} (\bibinfo{year}{1970}).

\bibitem[{\citenamefont{Kas and Rehr}(2017)}]{KasRehr17}
\bibinfo{author}{\bibfnamefont{J.~J.} \bibnamefont{Kas}} \bibnamefont{and}
  \bibinfo{author}{\bibfnamefont{J.~J.} \bibnamefont{Rehr}},
  \bibinfo{journal}{Phys. Rev. Lett.} \textbf{\bibinfo{volume}{119}},
  \bibinfo{pages}{176403} (\bibinfo{year}{2017}).

\bibitem[{\citenamefont{Vila et~al.}(2020)\citenamefont{Vila, Rehr, Kas, Peng,
  and Kowalski}}]{VilaRehr}
\bibinfo{author}{\bibfnamefont{F.~D.} \bibnamefont{Vila}},
  \bibinfo{author}{\bibfnamefont{J.~J.} \bibnamefont{Rehr}},
  \bibinfo{author}{\bibfnamefont{J.~J.} \bibnamefont{Kas}},
  \bibinfo{author}{\bibfnamefont{B.}~\bibnamefont{Peng}}, \bibnamefont{and}
  \bibinfo{author}{\bibfnamefont{K.}~\bibnamefont{Kowalski}},
  \bibinfo{journal}{UW Preprint}  (\bibinfo{year}{2020}).

\end{thebibliography}

\end{document}